\documentclass[aps,prl,reprint,superscriptaddress,showpacs]{revtex4-1}
\usepackage{graphicx}
\usepackage{textcomp}
\usepackage{gensymb}
\usepackage{xfrac}

\begin{document}

\title{Double-Q spin-density wave in iron arsenide superconductors}

\author{J. M. Allred}
\email{jallred@anl.gov}
\affiliation{Materials Science Division, Argonne National Laboratory, Argonne, IL 60439-4845, USA}
\author{K. M. Taddei}
\affiliation{Materials Science Division, Argonne National Laboratory, Argonne, IL 60439-4845, USA}
\affiliation{Physics Department, Northern Illinois University, DeKalb, IL 60115, USA}
\author{D. E. Bugaris}
\affiliation{Materials Science Division, Argonne National Laboratory, Argonne, IL 60439-4845, USA}
\author{M. J. Krogstad}
\affiliation{Materials Science Division, Argonne National Laboratory, Argonne, IL 60439-4845, USA}
\affiliation{Physics Department, Northern Illinois University, DeKalb, IL 60115, USA}
\author{S. H. Lapidus}
\affiliation{Advanced Photon Source, Argonne National Laboratory, Argonne, IL 60439-4845, USA}
\author{D. Y. Chung}
\author{H. Claus}
\affiliation{Materials Science Division, Argonne National Laboratory, Argonne, IL 60439-4845, USA}
\author{M. G. Kanatzidis}
\affiliation{Materials Science Division, Argonne National Laboratory, Argonne, IL 60439-4845, USA}
\affiliation{Department of Chemistry, Northwestern University, Evanston, IL 60208-3113, USA}
\author{D. E. Brown}
\affiliation{Physics Department, Northern Illinois University, DeKalb, IL 60115, USA}
\author{J. Kang}
\author{R. M. Fernandes}
\affiliation{School of Physics and Astronomy, University of Minnesota, Minneapolis, MN 55455, USA}
\author{I. Eremin}
\affiliation{Institut f\"ur Theoretische Physik III, Ruhr-Universit\"at Bochum, 44801 Bochum, Germany}
\author{S. Rosenkranz}
\affiliation{Materials Science Division, Argonne National Laboratory, Argonne, IL 60439-4845, USA}
\author{O. Chmaissem}
\affiliation{Materials Science Division, Argonne National Laboratory, Argonne, IL 60439-4845, USA}
\affiliation{Physics Department, Northern Illinois University, DeKalb, IL 60115, USA}
\author{R. Osborn}
\affiliation{Materials Science Division, Argonne National Laboratory, Argonne, IL 60439-4845, USA}

\date{\today}

\begin{abstract}
Elucidating the nature of the magnetic ground state of iron-based
superconductors is of paramount importance in unveiling the mechanism behind
their high temperature superconductivity. Until recently, it was thought that
superconductivity emerges only from an orthorhombic antiferromagnetic stripe
phase, which can in principle be described in terms of either localized or
itinerant spins. However, we recently reported that tetragonal symmetry is
restored inside the magnetically ordered state of a hole-doped
BaFe$_2$As$_2$. This observation was interpreted as indirect evidence of a
new double-\textbf{Q} magnetic structure, but alternative models of orbital
order could not be ruled out. Here, we present M\"ossbauer data that show
unambiguously that half of the iron sites in this tetragonal phase are
non-magnetic, establishing conclusively the existence of a novel magnetic
ground state with a non-uniform magnetization that is inconsistent with
localized spins. We show that this state is naturally explained as the
interference between two spin-density waves, demonstrating the itinerant
character of the magnetism of these materials and the primary role played by
magnetic over orbital degrees of freedom.
\end{abstract}


\maketitle

\section{Introduction}
One of the central questions to be answered in the iron-based superconductors
is the nature of their magnetic interactions. Because superconductivity
occurs in proximity to a magnetic instability, it is believed that magnetic
fluctuations play a key role in promoting superconducting order
\cite{Chubukov_review,Hirschfeld_review}. In these materials, the iron atoms
in each plane sit on a square lattice and the antiferromagnetic state, from
which superconductivity emerges, usually consists of stripes of iron spins
aligned ferromagnetically along one iron-iron bond direction and
antiferromagnetically along the the other, with a two-fold symmetry that
breaks the four-fold symmetry of the high temperature phase. Different
theoretical approaches have been proposed to describe the origin of this
magnetic two-fold ($C_2$) state, as well as the associated ``nematic" state,
and its relationship to superconductivity
\cite{Dai:2012em,Yin:2011ge,Yu:2013xx,Fernandes:2014jf}.

On the one hand, the large resistivities and enhanced effective masses of
the iron arsenides and chalcogenides have been interpreted as evidence for
proximity to a Mott transition, as seen in the similar phase diagrams of
cuprate superconductors \cite{Si08,JPHu08,Medici14}. This favours an approach
based on localized spin models, in which the iron spins $\mathbf{S}_{i}$,
with fixed amplitude $M$, live on the sites $i$ of the iron lattice and
interact with each other \textit{via} exchange interactions. This can give
rise to superconductivity with extended $s$-wave symmetry, in which the order
parameter changes sign between next-nearest neighbor sites. Some localized
models focus not on magnetic, but on orbital degrees of freedom, whose
fluctuations in general favour a regular $s$-wave state
\cite{Kruger:2009jm,Kontani:2012ii}. In this case, magnetic order is a
secondary effect of the four-fold symmetry breaking produced by changing the
relative occupation of the $d_{xz}$ and $d_{yz}$ iron orbitals.

On the other hand, itinerant spin models rely on the metallic character of
these compounds and on quasi-nesting features of their Fermi surfaces
\cite{Chubukov:2008p13399,Fernandes:2010p32347}. In this case, instead of
local spins on the lattice sites, the magnetism is best described as a
modulation of the spin polarization of the itinerant electrons,
\textit{i.e.}, a spin-density wave,
$\mathbf{S}\left(\mathbf{r}\right)=\mathbf{M}\cos\left(\mathbf{Q}\cdot\mathbf
{r}\right)$, with $\mathbf{Q}=\left(\pi,0\right)$ or $\left(0,\pi\right)$.
The resulting superconducting symmetry depends on details of the Fermi
surface, and is usually extended $s$-wave but can also be $d$-wave.
Determining which approach is valid, itinerant or localized, will therefore
have profound consequences both for the nature of the emergent
superconductivity and its relation to cuprate
superconductivity\cite{Fernandes:2014jf}.

Because both the localized and itinerant scenarios predict ground states with
the same space group, distinguishing between them is a challenging task.
Recently, we observed a new magnetic phase in hole-doped BaFe$_2$As$_2$ that
offers a new way to resolve this issue\cite{Avci:2014fp}. In pure
BaFe$_2$As$_2$, the structural, orbital and magnetic transitions occur
simultaneously in a first-order transition. When sodium is doped onto the
barium sites, the transition temperature, $T_N$, is reduced until magnetic
order is destroyed at $x\lesssim 0.3$\cite{Avci:2013iua}. However, between
$0.24\lesssim x \lesssim 0.3$, four-fold ($C_4$) symmetry is restored inside
the magnetically ordered state at $T_{\rm{r}}$ ($T_{\rm{r}}<T_{N}$), with a
reorientation of the magnetic moments along the $c$-axis
\cite{Wasser:2015fw}. Because the magnetic Bragg peaks have the same
reciprocal lattice indices above and below $T_{r}$, this $C_4$ phase was
interpreted as a double-\textbf{Q} magnetic structure described by a coherent
superposition of two spin-density waves,
$\mathbf{S}\left(\mathbf{r}\right)=\mathbf{M}_{1}\cos\left(\mathbf{Q}_{1}
\cdot\mathbf{r}\right)+\mathbf{M}_{2}\cos\left(\mathbf{Q}_{2}\cdot\mathbf{r}
\right)$, with $\mathbf{Q}_1=\left(\pi,0\right)$ and $\mathbf{Q}_2=
\left(0,\pi\right)$.

Although such order is naturally predicted by itinerant models for the
iron-based superconductors
\cite{Lorenzana11,Eremin11,Avci:2014fp,Brydon11,Wang2014xx}, it was not
possible to rule out the existence of an orbitally ordered structure that is
consistent with the observed $C_4$ phase. Furthermore, the existence of
magnetic Bragg peaks at both $\mathbf{Q}_1$ and $\mathbf{Q}_2$ could be
ascribed to domains of stripe magnetic order instead of double-\textbf{Q}
order. Khalyavin \textit{et al.}\ discuss a number of orbitally ordered
structures that are compatible with the diffraction
\cite{Khalyavin:2014dg}. Common to all of them is that the crystal structure
is tetragonal, but the magnetic structure still consists of orthorhombic
(\textit{i.e.}, single-\textbf{Q}) stripes.

In this paper, we present unambiguous evidence that the magnetic state of a
hole-doped iron arsenide, Sr$_{1-x}$Na$_x$Fe$_2$As$_2$ with $x=0.37$, is a
double-\textbf{Q} spin-density wave. This is the third hole-doped compound
within the $A$Fe$_2$As$_2$ series ($A=$ Ba, Sr) to show signatures of the
$C_4$ phase, indicating that this is a universal feature of hole-doping
\cite{Avci:2014fp,Bohmer:2014uk,Allred:2015tl}. We were able to synthesize
powders that exhibit a complete transformation of the sample to the $C_4$
phase below $T_{\rm{r}}$, allowing us to utilize M\"ossbauer spectroscopy as
a local probe of the magnetization on the $^{57}$Fe sites. Below
$T_{\rm{N}}$, within the $C_2$ phase, all sites show the same Zeeman
splitting due to internal molecular fields as expected for the
single-\textbf{Q} stripes. However, below the $C_4$ transition at
$T_{\rm{r}}<T_{\rm{N}}$, 50\% of the iron sites are non-magnetic while the
other 50\% show a doubling of the magnetization, exactly as expected from a
double-\textbf{Q} structure formed by the interference between two collinear
spin-density waves,
$\mathbf{S}\left(\mathbf{r}\right)=\mathbf{M}_{1}\cos\left(\mathbf{Q}_{1}
\cdot\mathbf{r}\right)+\mathbf{M}_{2}\cos\left(\mathbf{Q}_{2}\cdot
\mathbf{r}\right)$, with $\mathbf{M}_1$ = $\mathbf{M}_2$. This 
redistribution of magnetization density is not compatible with the idea of
local moments living on the iron sites or with any form of orbital order.
Therefore, our results point to the primary role played by itinerant
magnetism in the phase diagram of the iron arsenides, offering a key insight
into the nature of the electronic state from which superconductivity emerges.

\section{Results}
\subsection{X-ray and Neutron Diffraction}

\begin{figure}
\includegraphics[width =  \columnwidth]{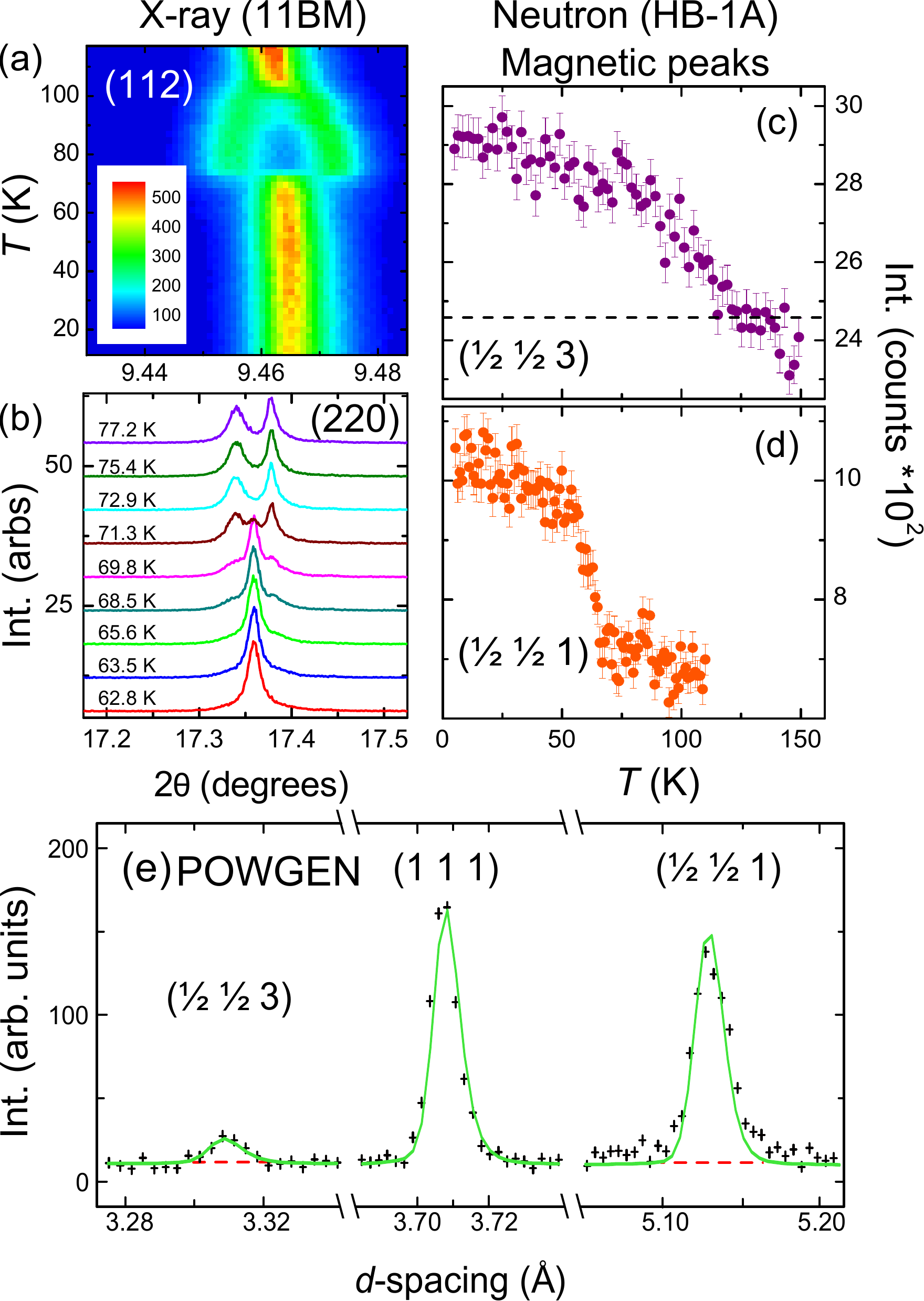}
\caption{
Temperature dependent diffraction data of Sr$_{0.63}$Na$_{0.37}$Fe$_2$As$_2$.
Left panels (a-b) are from x-rays (11BM, $\lambda = 0.413842$ \r{A}) and the
right panels (c-d) are from constant wavelength neutrons (HB-1A, $\lambda =
2.3626$ \r{A}). (a) is a false colour map of the data around the tetragonal
(112) peak, and (b) is a view of the tetragonal (220) peak at temperatures
near $T_{\rm{r}}$ taken from the same data. (c) and (d) show the intensity of
the ($\frac{1}{2}\frac{1}{2}3$) and ($\frac{1}{2}\frac{1}{2}1$) magnetic
peaks, respectively. (e) Detailed view of the calculated intensity (green
line) from the magnetic model fit to the 10 K POWGEN data (black crosses).
The dotted red line shows the calculated intensities of a non-magnetic model.
\label{Figure1}}
\end{figure}

Magnetism in the hole-doped series, Sr$_{1-x}$Na$_{x}$Fe$_2$As$_2$, has
higher transition temperatures and persists to higher levels of sodium
concentration than the equivalent barium series
\cite{CortesGil:2011ii,Avci:2013iua}.  We synthesized a compound with the
nominal composition of Sr$_{0.63}$Na$_{0.37}$Fe$_2$As$_2$, below the critical
phase boundary for magnetic order, which has a superconducting transition at
12\,K. Rietveld refinements of the x-ray powder diffraction spectra yielded a
sodium concentration of $x$ = 0.3691(5). More details of the sample
characterization are provided in the Methods section and the Supplementary
Information.

The transition from tetragonal ($I4/mmm$) to orthorhombic ($Fmmm$) symmetry
is evident in the powder x-ray diffraction as a splitting of some of the
Bragg peaks, such as the (112) peak, shown in Figure \ref{Figure1}a.  This
$C_2$ transition, which is either weakly first order or second order, occurs
around $T_{\rm{N}}\approx105$(2)\,K, below which the orthorhombic order parameter
(\textit{i.e.}, the magnitude of peak splitting) increases rapidly with
decreasing temperature. This behaviour is similar to many other iron-based
superconductors, but more unusually, this sample then transforms back to
tetragonal symmetry at a strongly first order transition at
$T_{\rm{r}}\approx 73\,K$. The first-order nature of this transition from
$C_2$ to $C_4$ symmetry can be seen in Figure \ref{Figure1}b, which shows
that the two phases coexist for $\sim 10$\,K below $T_{\rm{r}}$.

Powder neutron diffraction confirms that both the $C_2$ and $C_4$ phases are
magnetically ordered. Both the ($\frac{1}{2}\frac{1}{2}1$) and
($\frac{1}{2}\frac{1}{2}3$) magnetic peaks are present below $T_{\rm{N}}$
(Figure \ref{Figure1}c and d), but the increase in intensity of the former at
$T_{\rm{r}}$ shows that there is a significant spin reorientation in the
$C_4$ phase, as observed in Ba$_{1-x}$Na$_x$Fe$_2$As$_2$ ($0.24 \leq x \leq
0.28$) \cite{Avci:2014fp}. Since the transformation back to tetragonal
symmetry is complete in Sr$_{0.63}$Na$_{0.37}$Fe$_2$As$_2$ below 60\,K, we
are able to obtain a more reliable refinement of the magnetic structure than
was possible in the earlier work.  Figure \ref{Figure1}e shows that the data
are fit well by a model with moments along $c$-axis, in agreement with the 
moment direction deduced by Wa{\ss}er \textit{et al.}\ \cite{Wasser:2015fw}.  

\begin{figure}
\includegraphics[width=\columnwidth]{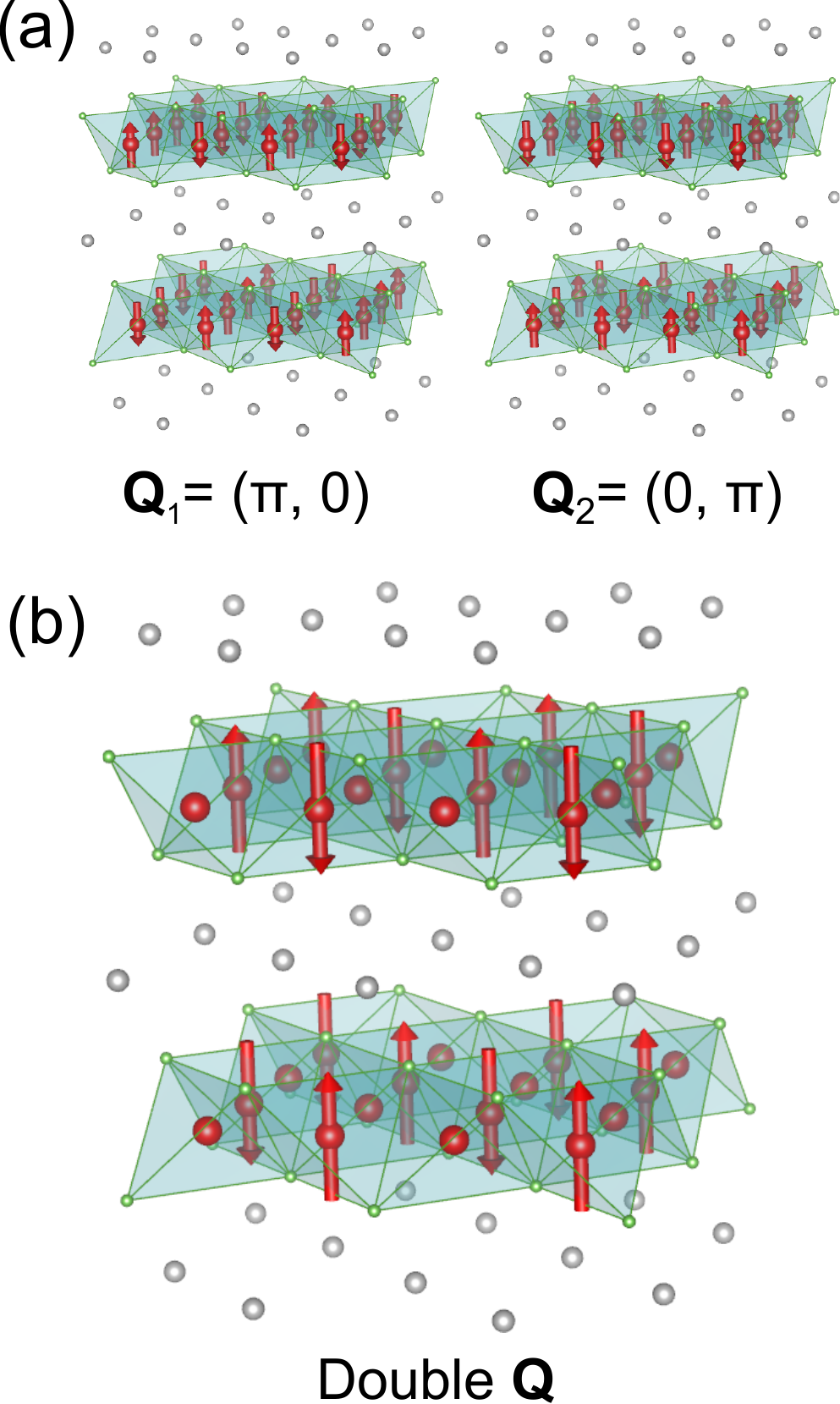}
\caption{
Single-\textbf{Q} and double-\textbf{Q} magnetic models.
(a) Single-\textbf{Q} model, in which spins are modulated with either
$\mathbf{Q}_{1}=\left(\pi,0\right)$ or $\mathbf{Q}_{2}=\left(0,\pi\right)$,
parallel to the $a$ and $b$ axes, respectively, in different domains. (b)
Double-\textbf{Q} model that is formed from the superposition of modulations
along $\mathbf{Q}_1$ and $\mathbf{Q}_2$.
\label{Figure2}}
\end{figure}

The magnetic Bragg peaks in the $C_4$ phase have the same reciprocal lattice
indices as the $C_2$ phase, using the tetragonal unit cell, so one possible
interpretation of the data is that the two phases have identical magnetic
stripe order, only differing by the orientation of the iron spins. In this
single-\textbf{Q} model, Bragg peaks from stripes parallel to the $x$ and $y$
axes in different domains would be incoherently superposed (Fig.
\ref{Figure2}a). Such a model would be magnetically orthorhombic, so
magnetoelastic coupling should generate an orthorhombic structural distortion
as well, but it is plausible that it is much weaker because of the spin
reorientation and therefore difficult to resolve.

An alternative interpretation is that there is a single domain comprising a
coherent superposition of magnetic stripes parallel to both the $x$ and $y$
axes, a double-\textbf{Q} model (Fig. \ref{Figure2}b). This is the model
predicted by itinerant approaches, in which magnetic order in the $C_4$ phase
is generated by band nesting along the $x$ and $y$ directions simultaneously,
restoring tetragonal symmetry \cite{Avci:2014fp}.  In such a case, the
residual spin-orbit coupling allows the parallel orientation of the resulting
magnetic moments from each wave vector only if they are along the
$z$-direction. It is well known that diffraction alone is unable to
distinguish between multi-domain single-\textbf{Q} and single-domain
multi-\textbf{Q} structures, since they produce identical Bragg peak
intensities. As discussed in Ref. \citenum{Khalyavin:2014dg}, it might be
possible to distinguish them with resonant x-ray scattering, which is
sensitive to the orbital configuration of the iron $d$-electrons. In
particular, space groups compatible with any possible orbital order would be
incompatible with a double-\textbf{Q} model.

\begin{figure}[!t]
\includegraphics[width=\columnwidth]{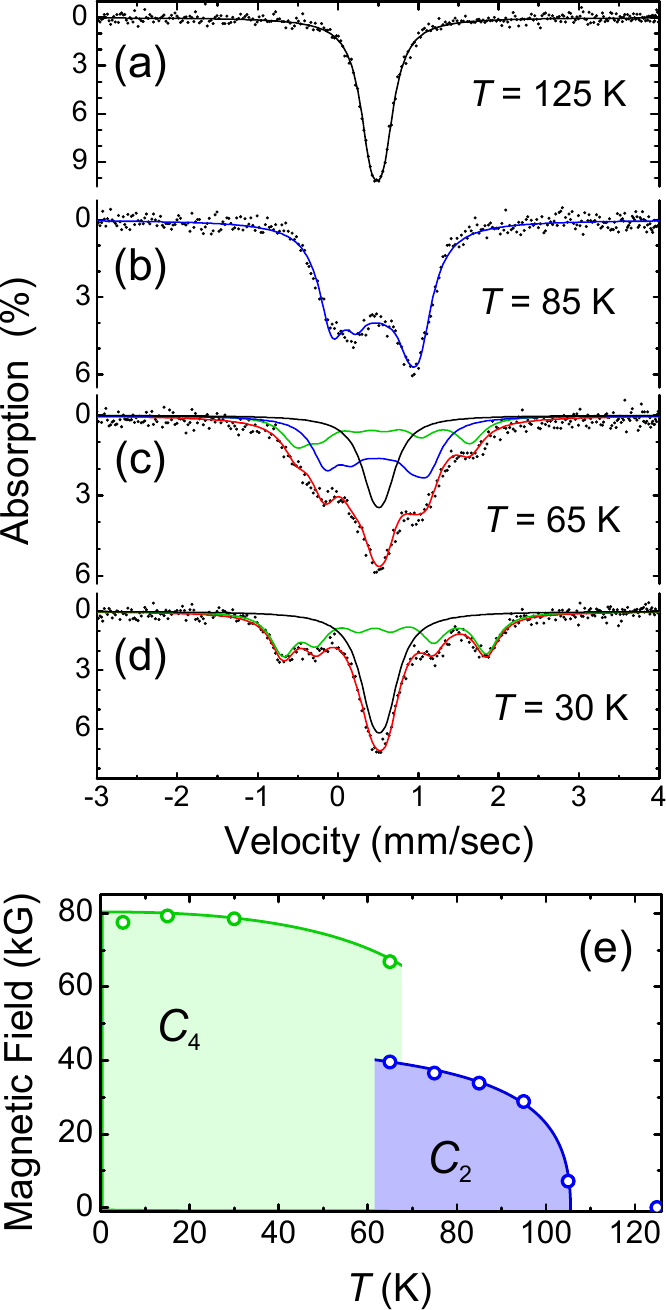}
\caption{
(a-d) M\"ossbauer spectra (black circles) at 
30, 65, 85, and 125\,K (5, 15, 75, 95, and 105\,K not shown). Fits to the
contributions of different iron sites are shown separately: non-magnetic
sites (black lines), $C_4$ magnetic sites (green lines), $C_2$ magnetic sites
(blue lines, total spectrum (red lines). (e) Effective magnetic field on each
magnetic site as a function of temperature determined from M\"ossbauer
spectroscopy. The lines are guides to the eye. Error bars are smaller than
the points.
\label{Figure3}}
\end{figure}

\subsection{M\"ossbauer Spectroscopy}

Although in reciprocal space, the single-\textbf{Q} and double-\textbf{Q}
models look identical, they are remarkably different in real space. As shown
in Figure \ref{Figure2}b, the coherent superposition of the orthogonal
stripes in the double-\textbf{Q} model results in a doubling of the magnetic
moments on half of the sites and a complete cancellation of the magnetic
moments on the other half. That is, half of the iron sites are spin-density
wave nodes. In local moment systems, nodes represent fluctuating spins that
have a high entropy, but in an itinerant spin-density wave, they can be a
natural consequence of a spatially inhomogeneous magnetization.

The best way to distinguish these two magnetic structures is therefore to use
a local probe of the magnetization. M\"{o}ssbauer spectroscopy is ideal
because the Zeeman splitting of the nuclear levels of $^{57}$Fe atoms is
directly proportional to the static magnetization density at the nuclear
site. Earlier M\"ossbauer spectra on iron arsenides were consistent with the
temperature dependence of the conventional antiferromagnetic stripe order, 
which is characterized by a single hyperfine field at each temperature 
\cite{McGuire:2008ji, Kitao:2008kk}.

We have measured M\"{o}ssbauer spectra at temperatures between 5\,K and
125\,K (Figure \ref{Figure3}). The 125\,K spectrum shows, as expected, a
single peak associated with the paramagnetic phase, with a small isomer shift
due to the chemical environment that is independent of temperature. Spectra
measured in the $C_2$ phase at 95, 85 and 75\,K (only 85\,K is shown) are
well fit with a single hyperfine field, characteristic of a single magnetic
site, which grows with decreasing temperature (Figure \ref{Figure3}e).
However, well below the $C_4$ transition, at 30\,K, there is a qualitatively
different spectrum, which consists of a large central peak with the same
isomer shift as the paramagnetic phase, indicating the presence of
non-magnetic sites, and a sextet indicating magnetic sites with a
significantly larger effective field than in the $C_2$ phase (by a factor of
$\sim2$).  A free fit to such a two-site model shows that the spectral
weights of each component are identical within the statistical uncertainty.
In other words, 50\% of the sites are magnetic and 50\% are non-magnetic,
exactly as predicted by the double-\textbf{Q} model.

The 65\,K spectrum, which was taken in the temperature range where diffraction
data indicated a co-existence of the $C_4$ and $C_2$ phases, shows
evidence of the superposition of three components,  two magnetic and one
non-magnetic. Although the parameters are too highly correlated to be fit
independently, the spectrum is consistent with a $C_4$ contribution,
comprising an equal concentration of large moment and non-magnetic sites, and
a $C_2$ contribution from smaller moment sites. Another parameter, the
electric field gradient, which is sensitive to the point-group symmetry of
the surrounding ions, changes sign between the $C_2$ and $C_4$ phases, but is
otherwise nearly temperature independent. The resulting local magnetization
of the magnetic sites as a function of temperature (Figure \ref{Figure3}e)
shows a clear doubling of the magnetic moment within the $C_4$ phase compared
to the $C_2$ phase, demonstrating that the $C_4$ magnetic structure involves
a redistribution of magnetization density from the non-magnetic to the
magnetic sites, an effect that is a clear fingerprint of an itinerant spin
density wave.

\section{Discussion and Conclusion\label{conclusion}}

The M\"{o}ssbauer data provide unambiguous evidence that the magnetism in the
$C_4$ phase is a double-\textbf{Q} spin-density wave. This has a number of
important consequences. As pointed out in Ref. \citenum{Khalyavin:2014dg},
such a double-\textbf{Q} magnetic structure is incompatible with either
ferro-orbital order involving the $d_{xz}$ and $d_{yz}$ iron orbitals or with
more complex patterns of orbital ordering. As a result, it implies that the
nematic phase observed in the underdoped compounds is not the cause, but a
consequence of magnetism, in agreement with the spin-nematic scenario
\cite{Fernandes:2012dv}. Although this observation by itself cannot rule out
the existence of orbital fluctuations, which favour the more conventional
$s$-wave superconducting state, it does indicate the primary role played by
magnetic fluctuations, which favour the unconventional sign-changing extended
$s$-wave state.

The most important conclusion to be drawn from this work is that the nature
of the $C_4$ magnetic state is not consistent with a model of localized spins
on the iron sites, in which every iron site is magnetic. At least within the
$t$-$J_1$-$J_2$ model, widely employed as an effective model to study
these materials \cite{JPHu08,Phillips09,Yu:2014fe}, such a non-uniform magnetization is not a
ground state of the model. It remains to be seen whether modifications of
this model, such as the inclusion of non-Heisenberg exchange interactions, like
the biquadratic or ring exchanges, could describe the non-uniform state.

By contrast, an itinerant approach offers a natural explanation of this
non-uniform magnetic structure as the interference of two nesting-related spin 
density waves,
$\mathbf{S}\left(\mathbf{r}\right)=\mathbf{M}_{1}\cos\left(\mathbf{Q}_{1}
\cdot\mathbf{r}\right)+\mathbf{M}_{2}\cos\left(\mathbf{Q}_{2}\cdot
\mathbf{r}\right)$, with $\mathbf{M}_1$ and $\mathbf{M}_2$ parallel to each
other. The fact that $M_1=M_2$ ensures not only the tetragonal symmetry of
the system, in agreement with the experimental observations, but also implies
that half of the sites are non-magnetic with their spin density transferred
to neighbouring sites with double the magnetization. This is a remarkable
observation that is only compatible with itinerant electrons. It is also
consistent with the prediction of itinerant models that such a state becomes
favoured over the stripe state for large enough doping levels
\cite{Lorenzana11,Brydon11,Avci:2014fp,Wang2014xx,Gastiasoro:2015vf}.
Furthermore, a secondary checkerboard charge order should accompany this
non-uniform phase, in which the non-magnetic sites have locally a different
charge density than the magnetic sites \cite{Fernandes15}. It has been argued
that fluctuations of this secondary charge order can enhance the extended
$s$-wave transition temperature \cite{Fernandes15}.

The reorientation of the magnetization along the $c$-axis follows from
general group-theory arguments related to the space-group of a single FeAs
plane with preserved tetragonal symmetry \cite{Cvetkovic:2013ge}. In the iron
pnictides, spin-orbit coupling is not small \cite{2014arXiv1409.8669B}, and
as a consequence, possible spin orientations are restricted to certain
crystallographic directions. In particular, a group-theory analysis reveals
three possibilities:
$\mathbf{M}_{1}\parallel\hat{\mathbf{x}}$
and $\mathbf{M}_{2}\parallel\hat{\mathbf{y}}$; 
$\mathbf{M}_{1}\parallel\hat{\mathbf{y}}$ and 
$\mathbf{M}_{2}\parallel\hat{\mathbf{x}}$; or 
$\mathbf{M}_{1}\parallel\hat{\mathbf{z}}$
and $\mathbf{M}_{2}\parallel\hat{\mathbf{z}}$. Because only the latter
is compatible with the non-uniform state discussed here, the spins
must point along the $c$-axis. This is discussed in more detail in the 
Supplementary Information.

We note that the itinerancy of the magnetism of the iron pnictides does not
imply that interactions are necessarily weak
\cite{Mazin09,Wei_Ku10,Bascones12}. Indeed, even in elemental iron, a
weak-coupling approach does not fully describe the properties of the
ferromagnetic state. In the iron pnictides, interaction effects were shown to
be important to capture high-energy properties of the spin spectrum
\cite{Yin_Kotliar11} and the sizable fluctuating moment observed in the
paramagnetic state \cite{Arita10}. Thus, it is likely that interactions are
moderate, and may affect distinct families of iron-based superconductors,
such as the iron chalcogenides, differently \cite{Dai:2012em,Yin:2011ge}. At
least in the iron pnictides, however, our work demonstrates that itinerancy
is an essential ingredient of these fascinating materials.

\section{Methods\label{Methods}}
\subsection{Synthesis}
Handling of all starting materials was performed in an M-Braun glovebox under
an inert Ar atmosphere ($<0.1$\,ppm of H$_2$O and O$_2$).  Sr (Aldrich,
99.9\%) and Fe (Alfa Aesar, 99.99+\%) were used as received.  Small pieces of
Na free of oxide coating were trimmed from large lumps (Aldrich, 99\%).
Granules of As (Alfa Aesar, 99.99999+\%) were ground to a coarse powder prior
to use. The precursor materials SrAs, NaAs, and Fe$_2$As were synthesized in
quartz tubes from stoichiometric reactions of the elements at 800$\celsius$,
350$\celsius$, and 700$\celsius$ respectively.  Polycrystalline samples of
Sr$_{0.67}$Na$_{0.37}$Fe$_2$As$_2$ were prepared from stoichiometric mixtures of
SrAs, NaAs, and Fe$_2$As, which were ground thoroughly with a mortar and
pestle, and loaded in alumina crucibles.  The alumina crucibles were sealed
in Nb tubes under Ar, which were further sealed in quartz tubes under vacuum.
The reaction mixtures were subjected to multiple heating cycles between
850-950$\celsius$ for durations less than 48\,h (to minimize loss of Na by
volatilization). The samples underwent grinding by mortar and pestle between
heating cycles in order to homogenize the compositions. Following the final
heating cycles, the sealed samples were quenched in air from the maximum
temperature rather than allowing them to cool slowly.  Initial
characterization of the dark gray powders was conducted by laboratory powder
X-ray diffraction and magnetization measurements. More details of the sample
characterization are provided in the Supplementary Information.

\subsection{Powder Diffraction}
X-ray powder diffraction measurements were performed at Argonne National
Laboratory using beamline 11-BM at the Advanced Photon Source. Neutron powder
diffraction measurements were performed at Oak Ridge National Laboratory
using beamline HB-1A at the High Flux Isotope Reactor and the POWGEN
diffractometer at the Spallation Neutron Source.

\subsection{M\"ossbauer Spectroscopy}
M\"ossbauer measurements were performed in transmission geometry with a
sinusoidally driven 2\,mCi $^{57}$Co(Rh) source and a germanium detector.
Silicon diode sensors allowed the control and stabilization of the sample
temperature to within 0.2 K for a conventional bath cryostat.  Powder samples
having an effective area density of 4\,mg/cm$^{2}$ of $^{57}$Fe were placed on
99.999\% pure aluminum foil held in place by kapton tape.  Calibrations were
made using a natural $\alpha$-Fe foil.  The spectra were fit by varying the
isomer shift, magnetic hyperfine field, and the electric quadrupole factor. 
The intensities of the magnetic sextet-split lines were constrained to a
1:2:3 ratio according to their Clebsch-Gordon coefficients (or magnetic
dipole matrix elements), and the Lorentzian linewidths for all lines of a
particular iron site were constrained to be the same. Details of the fit
parameters are given in the Supplementary Information.

\bibliography{Sr122-Na37-manuscript}

\begin{thebibliography}{10}
\expandafter\ifx\csname url\endcsname\relax
  \def\url#1{\texttt{#1}}\fi
\expandafter\ifx\csname urlprefix\endcsname\relax\def\urlprefix{URL }\fi
\providecommand{\bibinfo}[2]{#2}
\providecommand{\eprint}[2][]{\url{#2}}

\bibitem{Chubukov_review}
\bibinfo{author}{Chubukov, A.}
\newblock \bibinfo{title}{Pairing mechanism in fe-based superconductors}.
\newblock \emph{\bibinfo{journal}{Ann. Rev. Cond. Matt. Phys.}}
  \textbf{\bibinfo{volume}{3}}, \bibinfo{pages}{57--92} (\bibinfo{year}{2012}).

\bibitem{Hirschfeld_review}
\bibinfo{author}{Hirschfeld, P.}, \bibinfo{author}{Korshunov, M.} \&
  \bibinfo{author}{Mazin, I.}
\newblock \bibinfo{title}{Gap symmetry and structure of fe-based
  superconductors}.
\newblock \emph{\bibinfo{journal}{Rep. Prog. Phys.}}
  \textbf{\bibinfo{volume}{74}}, \bibinfo{pages}{124508}
  (\bibinfo{year}{2011}).

\bibitem{Dai:2012em}
\bibinfo{author}{Dai, P.}, \bibinfo{author}{Hu, J.} \&
  \bibinfo{author}{Dagotto, E.}
\newblock \bibinfo{title}{{Magnetism and its microscopic origin in iron-based
  high-temperature superconductors}}.
\newblock \emph{\bibinfo{journal}{Nature Phys.}} \textbf{\bibinfo{volume}{8}},
  \bibinfo{pages}{709--718} (\bibinfo{year}{2012}).

\bibitem{Yin:2011ge}
\bibinfo{author}{Yin, Z.~P.}, \bibinfo{author}{Haule, K.} \&
  \bibinfo{author}{Kotliar, G.}
\newblock \bibinfo{title}{{Kinetic frustration and the nature of the magnetic
  and paramagnetic states in iron pnictides and iron~chalcogenides}}.
\newblock \emph{\bibinfo{journal}{Nature Mater.}}
  \textbf{\bibinfo{volume}{10}}, \bibinfo{pages}{932--935}
  (\bibinfo{year}{2011}).

\bibitem{Yu:2013xx}
\bibinfo{author}{Yu, R.}, \bibinfo{author}{Goswami, P.}, \bibinfo{author}{Si,
  Q.}, \bibinfo{author}{Nikolic, P.} \& \bibinfo{author}{Zhu, J.-X.}
\newblock \bibinfo{title}{{Superconductivity at the border of electron
  localization and itinerancy}}.
\newblock \emph{\bibinfo{journal}{Nature Comm.}} \textbf{\bibinfo{volume}{4}},
  \bibinfo{pages}{2783} (\bibinfo{year}{2013}).

\bibitem{Fernandes:2014jf}
\bibinfo{author}{Fernandes, R.~M.}, \bibinfo{author}{Chubukov, A.~V.} \&
  \bibinfo{author}{Schmalian, J.}
\newblock \bibinfo{title}{{What drives nematic order in iron-based
  superconductors?}}
\newblock \emph{\bibinfo{journal}{Nature Phys.}} \textbf{\bibinfo{volume}{10}},
  \bibinfo{pages}{97--104} (\bibinfo{year}{2014}).

\bibitem{Si08}
\bibinfo{author}{Si, Q.} \& \bibinfo{author}{Abrahams, E.}
\newblock \bibinfo{title}{Strong correlations and magnetic frustration in the
  high ${T}_{c}$ iron pnictides}.
\newblock \emph{\bibinfo{journal}{Phys. Rev. Lett.}}
  \textbf{\bibinfo{volume}{101}}, \bibinfo{pages}{076401}
  (\bibinfo{year}{2008}).

\bibitem{JPHu08}
\bibinfo{author}{Seo, K.}, \bibinfo{author}{Bernevig, B.~A.} \&
  \bibinfo{author}{Hu, J.}
\newblock \bibinfo{title}{Pairing symmetry in a two-orbital exchange coupling
  model of oxypnictides}.
\newblock \emph{\bibinfo{journal}{Phys. Rev. Lett.}}
  \textbf{\bibinfo{volume}{101}}, \bibinfo{pages}{206404}
  (\bibinfo{year}{2008}).

\bibitem{Medici14}
\bibinfo{author}{de' Medici, L.}, \bibinfo{author}{Giovannetti, G.} \&
  \bibinfo{author}{Capone, M.}
\newblock \bibinfo{title}{Selective mott physics as a key to iron
  superconductors}.
\newblock \emph{\bibinfo{journal}{Phys. Rev. Lett.}}
  \textbf{\bibinfo{volume}{112}}, \bibinfo{pages}{177001}
  (\bibinfo{year}{2014}).

\bibitem{Kruger:2009jm}
\bibinfo{author}{Kr{\"u}ger, F.}, \bibinfo{author}{Kumar, S.},
  \bibinfo{author}{Zaanen, J.} \& \bibinfo{author}{van~den Brink, J.}
\newblock \bibinfo{title}{{Spin-orbital frustrations and anomalous metallic
  state in iron-pnictide superconductors}}.
\newblock \emph{\bibinfo{journal}{Phys. Rev. B}} \textbf{\bibinfo{volume}{79}},
  \bibinfo{pages}{054504} (\bibinfo{year}{2009}).

\bibitem{Kontani:2012ii}
\bibinfo{author}{Kontani, H.}, \bibinfo{author}{Inoue, Y.},
  \bibinfo{author}{Saito, T.}, \bibinfo{author}{Yamakawa, Y.} \&
  \bibinfo{author}{Onari, S.}
\newblock \bibinfo{title}{{Orbital fluctuation theory in iron-based
  superconductors: s$^{++}$ -wave superconductivity, structure transition, and
  impurity-induced nematic order}}.
\newblock \emph{\bibinfo{journal}{Sol. Stat. Comm.}}
  \textbf{\bibinfo{volume}{152}}, \bibinfo{pages}{718--727}
  (\bibinfo{year}{2012}).

\bibitem{Chubukov:2008p13399}
\bibinfo{author}{Chubukov, A.~V.}, \bibinfo{author}{Efremov, D.~V.} \&
  \bibinfo{author}{Eremin, I.}
\newblock \bibinfo{title}{{Magnetism, superconductivity, and pairing symmetry
  in iron-based superconductors}}.
\newblock \emph{\bibinfo{journal}{Phys. Rev. B}} \textbf{\bibinfo{volume}{78}},
  \bibinfo{pages}{134512} (\bibinfo{year}{2008}).

\bibitem{Fernandes:2010p32347}
\bibinfo{author}{Fernandes, R.~M.} \emph{et~al.}
\newblock \bibinfo{title}{{Unconventional pairing in the iron arsenide
  superconductors}}.
\newblock \emph{\bibinfo{journal}{Phys. Rev. B}} \textbf{\bibinfo{volume}{81}},
  \bibinfo{pages}{140501} (\bibinfo{year}{2010}).

\bibitem{Avci:2014fp}
\bibinfo{author}{Avci, S.} \emph{et~al.}
\newblock \bibinfo{title}{{Magnetically driven suppression of nematic order in
  an iron-based superconductor}}.
\newblock \emph{\bibinfo{journal}{Nature Comm.}} \textbf{\bibinfo{volume}{5}},
  \bibinfo{pages}{3845} (\bibinfo{year}{2014}).

\bibitem{Avci:2013iua}
\bibinfo{author}{Avci, S.} \emph{et~al.}
\newblock \bibinfo{title}{{Structural, magnetic, and superconducting properties
  of Ba$_{1-x}$Na$_{x}$Fe$_{2}$As$_{2}$}}.
\newblock \emph{\bibinfo{journal}{Phys. Rev. B}} \textbf{\bibinfo{volume}{88}},
  \bibinfo{pages}{094510} (\bibinfo{year}{2013}).

\bibitem{Wasser:2015fw}
\bibinfo{author}{Wa{\ss}er, F.} \emph{et~al.}
\newblock \bibinfo{title}{{Spin reorientation in
  Ba$_{0.65}$Na$_{0.35}$Fe$_{2}$As$_{2}$ studied by single-crystal neutron
  diffraction}}.
\newblock \emph{\bibinfo{journal}{Phys. Rev. B}} \textbf{\bibinfo{volume}{91}},
  \bibinfo{pages}{060505} (\bibinfo{year}{2015}).

\bibitem{Lorenzana11}
\bibinfo{author}{Giovannetti, G.} \emph{et~al.}
\newblock \bibinfo{title}{Proximity of iron pnictide superconductors to a
  quantum tricritical point}.
\newblock \emph{\bibinfo{journal}{Nature Comm.}} \textbf{\bibinfo{volume}{2}},
  \bibinfo{pages}{398} (\bibinfo{year}{2011}).

\bibitem{Eremin11}
\bibinfo{author}{Eremin, I.} \& \bibinfo{author}{Chubukov, A.~V.}
\newblock \bibinfo{title}{Magnetic degeneracy and hidden metallicity of the
  spin-density-wave state in ferropnictides}.
\newblock \emph{\bibinfo{journal}{Phys. Rev. B}} \textbf{\bibinfo{volume}{81}},
  \bibinfo{pages}{024511} (\bibinfo{year}{2010}).

\bibitem{Brydon11}
\bibinfo{author}{Brydon, P. M.~R.}, \bibinfo{author}{Schmiedt, J.} \&
  \bibinfo{author}{Timm, C.}
\newblock \bibinfo{title}{Microscopically derived ginzburg-landau theory for
  magnetic order in the iron pnictides}.
\newblock \emph{\bibinfo{journal}{Phys. Rev. B}} \textbf{\bibinfo{volume}{84}},
  \bibinfo{pages}{214510} (\bibinfo{year}{2011}).

\bibitem{Wang2014xx}
\bibinfo{author}{Wang, X.}, \bibinfo{author}{Kang, J.} \&
  \bibinfo{author}{Fernandes, R.~M.}
\newblock \bibinfo{title}{Magnetic order without tetragonal-symmetry-breaking
  in iron arsenides: Microscopic mechanism and spin-wave spectrum}.
\newblock \emph{\bibinfo{journal}{Phys. Rev. B}} \textbf{\bibinfo{volume}{91}},
  \bibinfo{pages}{024401} (\bibinfo{year}{2015}).

\bibitem{Khalyavin:2014dg}
\bibinfo{author}{Khalyavin, D.~D.} \emph{et~al.}
\newblock \bibinfo{title}{{Symmetry of reentrant tetragonal phase in
  Ba$_{1-x}$Na$_x$Fe$_2$As$_2$ : Magnetic versus orbital ordering mechanism}}.
\newblock \emph{\bibinfo{journal}{Phys. Rev. B}} \textbf{\bibinfo{volume}{90}},
  \bibinfo{pages}{174511} (\bibinfo{year}{2014}).

\bibitem{Bohmer:2014uk}
\bibinfo{author}{B{\"o}hmer, A.~E.} \emph{et~al.}
\newblock \bibinfo{title}{{Superconductivity-induced reentrance of orthorhombic
  distortion in Ba$_{1-x}$K$_x$Fe$_2$As$_2$}}.
\newblock \emph{\bibinfo{journal}{arXiv}} \bibinfo{pages}{1412.7038v2}
  (\bibinfo{year}{2014}).

\bibitem{Allred:2015tl}
\bibinfo{author}{Allred, J.~M.} \emph{et~al.}
\newblock \bibinfo{title}{{Tetragonal magnetic phase in
  Ba$_{1-x}$K$_{x}$Fe$_{2}$As$_{2}$ from x-ray and neutron diffraction}}.
\newblock \emph{\bibinfo{journal}{arXiv}} \bibinfo{pages}{1505.01433v1}
  (\bibinfo{year}{2015}).

\bibitem{CortesGil:2011ii}
\bibinfo{author}{Cortes-Gil, R.} \& \bibinfo{author}{Clarke, S.~J.}
\newblock \bibinfo{title}{{Structure, Magnetism, and Superconductivity of the
  Layered Iron Arsenides Sr$_{1-x}$Na$_{x}$Fe$_{2}$As$_{2}$}}.
\newblock \emph{\bibinfo{journal}{Chem. Mater.}} \textbf{\bibinfo{volume}{23}},
  \bibinfo{pages}{1009--1016} (\bibinfo{year}{2011}).

\bibitem{McGuire:2008ji}
\bibinfo{author}{McGuire, M.} \emph{et~al.}
\newblock \bibinfo{title}{{Phase transitions in LaFeAsO: Structural, magnetic,
  elastic, and transport properties, heat capacity and M{\"o}ssbauer spectra}}.
\newblock \emph{\bibinfo{journal}{Phys. Rev. B}} \textbf{\bibinfo{volume}{78}},
  \bibinfo{pages}{094517} (\bibinfo{year}{2008}).

\bibitem{Kitao:2008kk}
\bibinfo{author}{Kitao, S.} \emph{et~al.}
\newblock \bibinfo{title}{{Spin Ordering in LaFeAsO and Its Suppression in
  Superconductor LaFeAsO 0.89F 0.11Probed by M{\"o}ssbauer Spectroscopy}}.
\newblock \emph{\bibinfo{journal}{J. Phys. Soc. Japan}}
  \textbf{\bibinfo{volume}{77}}, \bibinfo{pages}{103706}
  (\bibinfo{year}{2008}).

\bibitem{Fernandes:2012dv}
\bibinfo{author}{Fernandes, R.~M.}, \bibinfo{author}{Chubukov, A.~V.},
  \bibinfo{author}{Knolle, J.}, \bibinfo{author}{Eremin, I.} \&
  \bibinfo{author}{Schmalian, J.}
\newblock \bibinfo{title}{{Preemptive nematic order, pseudogap, and orbital
  order in the iron pnictides}}.
\newblock \emph{\bibinfo{journal}{Phys. Rev. B}} \textbf{\bibinfo{volume}{85}},
  \bibinfo{pages}{024534} (\bibinfo{year}{2012}).

\bibitem{Phillips09}
\bibinfo{author}{Lv, W.}, \bibinfo{author}{Wu, J.} \&
  \bibinfo{author}{Phillips, P.}
\newblock \bibinfo{title}{Orbital ordering induces structural phase transition
  and the resistivity anomaly in iron pnictides}.
\newblock \emph{\bibinfo{journal}{Phys. Rev. B}} \textbf{\bibinfo{volume}{80}},
  \bibinfo{pages}{224506} (\bibinfo{year}{2009}).

\bibitem{Yu:2014fe}
\bibinfo{author}{Yu, R.}, \bibinfo{author}{Zhu, J.-X.} \& \bibinfo{author}{Si,
  Q.}
\newblock \bibinfo{title}{{Orbital-selective superconductivity, gap anisotropy,
  and spin resonance excitations in a multiorbital t-$J_1$-$J_2$ model for iron
  pnictides}}.
\newblock \emph{\bibinfo{journal}{Phys. Rev. B}} \textbf{\bibinfo{volume}{89}},
  \bibinfo{pages}{024509} (\bibinfo{year}{2014}).

\bibitem{Gastiasoro:2015vf}
\bibinfo{author}{Gastiasoro, M.~N.} \& \bibinfo{author}{Andersen, B.~M.}
\newblock \bibinfo{title}{{Competing magnetic double-Q phases and
  superconductivity-induced re-entrance of $C_2$ magnetic stripe order in iron
  pnictides}}.
\newblock \emph{\bibinfo{journal}{arXiv}} \bibinfo{pages}{1502.05859}
  (\bibinfo{year}{2015}).

\bibitem{Fernandes15}
\bibinfo{author}{{Fernandes}, R.~M.}, \bibinfo{author}{{Kivelson}, S.~A.} \&
  \bibinfo{author}{{Berg}, E.}
\newblock \bibinfo{title}{{Is there a hidden chiral density-wave in the
  iron-based superconductors?}}
\newblock \emph{\bibinfo{journal}{arXiv}} \bibinfo{pages}{1504.03656}
  (\bibinfo{year}{2015}).

\bibitem{Cvetkovic:2013ge}
\bibinfo{author}{Cvetkovic, V.} \& \bibinfo{author}{Vafek, O.}
\newblock \bibinfo{title}{{Space group symmetry, spin-orbit coupling, and the
  low-energy effective Hamiltonian for iron-based superconductors}}.
\newblock \emph{\bibinfo{journal}{Phys. Rev. B}} \textbf{\bibinfo{volume}{88}},
  \bibinfo{pages}{134510} (\bibinfo{year}{2013}).

\bibitem{2014arXiv1409.8669B}
\bibinfo{author}{Borisenko, S.} \emph{et~al.}
\newblock \bibinfo{title}{{Direct observation of spin-orbit coupling in
  iron-based superconductors}}.
\newblock \emph{\bibinfo{journal}{arXiv}} \bibinfo{pages}{1409.8669}
  (\bibinfo{year}{2014}).

\bibitem{Mazin09}
\bibinfo{author}{Johannes, M.~D.} \& \bibinfo{author}{Mazin, I.~I.}
\newblock \bibinfo{title}{Microscopic origin of magnetism and magnetic
  interactions in ferropnictides}.
\newblock \emph{\bibinfo{journal}{Phys. Rev. B}} \textbf{\bibinfo{volume}{79}},
  \bibinfo{pages}{220510} (\bibinfo{year}{2009}).

\bibitem{Wei_Ku10}
\bibinfo{author}{Yin, W.-G.}, \bibinfo{author}{Lee, C.-C.} \&
  \bibinfo{author}{Ku, W.}
\newblock \bibinfo{title}{Unified picture for magnetic correlations in
  iron-based superconductors}.
\newblock \emph{\bibinfo{journal}{Phys. Rev. Lett.}}
  \textbf{\bibinfo{volume}{105}}, \bibinfo{pages}{107004}
  (\bibinfo{year}{2010}).

\bibitem{Bascones12}
\bibinfo{author}{Bascones, E.}, \bibinfo{author}{Valenzuela, B.} \&
  \bibinfo{author}{Calder\'on, M.~J.}
\newblock \bibinfo{title}{Orbital differentiation and the role of orbital
  ordering in the magnetic state of fe superconductors}.
\newblock \emph{\bibinfo{journal}{Phys. Rev. B}} \textbf{\bibinfo{volume}{86}},
  \bibinfo{pages}{174508} (\bibinfo{year}{2012}).

\bibitem{Yin_Kotliar11}
\bibinfo{author}{Yin, Z.}, \bibinfo{author}{Haule, K.} \&
  \bibinfo{author}{Kotliar, G.}
\newblock \bibinfo{title}{Magnetism and charge dynamics in iron pnictides}.
\newblock \emph{\bibinfo{journal}{Nature Phys.}} \textbf{\bibinfo{volume}{7}},
  \bibinfo{pages}{294--297} (\bibinfo{year}{2011}).

\bibitem{Arita10}
\bibinfo{author}{Hansmann, P.} \emph{et~al.}
\newblock \bibinfo{title}{Dichotomy between large local and small ordered
  magnetic moments in iron-based superconductors}.
\newblock \emph{\bibinfo{journal}{Phys. Rev. Lett.}}
  \textbf{\bibinfo{volume}{104}}, \bibinfo{pages}{197002}
  (\bibinfo{year}{2010}).

\end{thebibliography}

\section{End Notes}
\subsection{Acknowledgements}
This work was supported by the U.S. Department of Energy, Office of Science,
Materials Sciences and Engineering Division and Scientific User Facilities
Division. X-ray experiments were performed at the Advanced Photon Source,
which is supported by the Office of Basic Energy Sciences under Contract No.
DE-AC02-06CH11357.  Neutron experiments were performed at the High Flux
Isotope Reactor and Spallation Neutron Source. R.M.F. and J.K. were supported by
the U.S. Department of Energy, Office of Science, Basic Energy Sciences,
under award number DE-SC0012336.The work of I.E. was supported by the Focus
Program 1458 Eisen-Pniktide of the DFG, and by the German Academic Exchange
Service (DAAD PPP USA no. 57051534). The authors thank A. A. Aczel, A. Huq,
M. J. Kirkham, and P. S. Whitfield for experimental assistance, E. E. Alp
for use of his M\"ossbauer spectrometer, and B. M. Andersen, A. V. Chubukov,
M. N. Gastiasoro, A. Yaresko, and Y. Zhao for fruitful discussions.

\subsection{Author contributions} 
Samples were prepared by D.E.Bu., with additional support from D.Y.C., and
M.G.K. The experiments were devised by J.M.A., K.M.T, O.C., S.R., and R.O.
The x-ray and neutron diffraction experiments  were performed by J.M.A,
K.M.T., O.C., M.J.K, S.R., and S.H.L.   M\"ossbauer spectroscopy was
performed by D.E.Br. Magnetization measurements were performed by H.C. The
data were analyzed by J.M.A, K.M.T., O.C., S.R., R.O, and D.E.Br.
Theoretical interpretation was provided by J.K., R.M.F. and I.E. The 
manuscript and supplementary information were written by J.M.A., R.O.,
R.M.F., and I.E. with input from all the authors.

\end{document}